\documentstyle[prl,aps]{revtex} 
\begin{document} 
\title{\Large Relativistic covariance and the bound state wave
function}  
\author{L.Micu\footnote{E-mail address: 
lmicu@theory.nipne.ro}} 
\address{Department of
Theoretical Physics\\ Horia Hulubei Institute for Physics and Nuclear
Engineering\\
 Bucharest POB MG-6, 76900 Romania}
\maketitle
\begin{abstract}
We establish a relation between the solution of a relativistic bound state
equation in quantum mechanics and the field representation of a bound state
with the aid of creation and annihilation operators. 
We show that a bound system can be represented by a gas of free constituents
and a classical effective field representing the countless quantum
fluctuations  generating the binding potential. The
distribution function of the internal momenta is given by the projection of
the free states on the solution of a relativistic bound state equation in the
rest frame of the bound system.  In this approach Lorentz covariance,
mass-shell constraints and single particle normalizability of the bound state
function are simultaneously and explicitly satisfied.
The discussion is made for a two particle bound state and can be easily
generalized to the case of three or more particles.
\end{abstract}  

\vskip1cm
The bound states of a two particle system found their best description in
nonrelativistic quantum mechanics where their existence is conditioned by the
presence of an attractive potential well. In this approach the internal
dynamics is represented by an unphysical particle with a reduced
mass, moving inside the well. The defining features of a bound state function
are stationarity and normalizability.

In the relativistic approaches derived from field theory the bound state wave
functions do not satisfy these criteria because the binding is supposed a
consequence of quantum fluctuations, that is of a continuous exchange of
quanta among the constituents \cite{bs,qp}.  The internal dynamics involves an
indefinite number of independent particles and the bound state wave function
has a  fluctuating character. The attempt to improve the situation by
restricting the intermediate states to the positive energy ones is artificial
and may destroy relativistic covariance \cite{qp}.  

It has been however shown by Dirac \cite{pamd} that it is possible to quantize
a bound system with a fixed number of particles and to satisfy in the same
time relativistic covariance by constructing explicitly the
interaction dependent generators of the symmetry group. Restraining the
symmetry group to those transformations which are purely kinematical it has
been possible to develop dynamical models independent on the concrete form of
the interaction. This idea stays at the origin of the light cone models
extensively used in hadron physics \cite{lc}. The full relativistic covariance
is of course destroyed in this case, but the effects of symmetry violation may
be evaluated by comparing various quantization schemes \cite{as}. 

Other stationary approaches like, for instance, those based on the relativistic
generalization of the time  independent Schr\"odinger equation with an
interaction potential yielding a realistic bound state mass spectrum \cite{gi}
have been used mainly in the study of static effects.

Closing this discussion we notice that the quantum
mechanical approach and the field approach can be put in agreement with 
each other if one finds a way to introduce the independent
constituents in the quantum mechanical approach and if one assumes that the
binding potential is the time averaged result of some fast quantum
fluctuations. The main problem remaining to be solved is then to find the
relation between the solution of a relativistic bound state equation and the
field representation of a bound state in terms of creation and annihilation
operators. The clarification of this point is the main purpose of the present
paper. 

In the following we refer to the meson case as a quark antiquark
bound state. The flavor and color indices are omitted for simplicity and
the interaction potential is assumed to be white. 

We first consider the bound state problem in the
meson rest frame. In order to ensure an independent treatment of the quarks
we assume that the total Hamiltonian is the sum of two free Dirac Hamiltonians
and of an interaction potential which depends on both coordinates. Its
eigenfunctions are the internal functions of the bound system in the rest
frame and the corresponding eigenvalues are the meson masses. We then have: 

\begin{equation}\label{h}
[\sum_{j=1,2}( -i\vec{\nabla}^{(j)}\vec{\alpha}^{(j)}+\beta m_j)+ {\mathcal
V}_0(\vec{x}_1,\vec{x}_2)]\Psi(\vec{x}_1,\vec{x}_2)=M
~\Psi(\vec{x}_1,\vec{x}_2) 
\end{equation}
where $\alpha$ and $\beta$ are Dirac matrices and ${\mathcal V}_0$ is the
interaction potential. The function $\Psi(\vec{x}_1,\vec{x}_2)$ has two
spinorial indices and can be best written under the form of a $4\times4$
matrix. 
 
We remark that the independent treatment of the quarks entails the
violation of translational invariance with respect to the quark
coordinates. In order to remove this inconsistency, we introduce a vector-type
external field $\vec{\mathcal V}(\vec{x}_1,\vec{x}_2)$ which enters the
following set of equations  

\begin{equation}\label{p}
 (-i\vec{\nabla}^{(1)}-i\vec{\nabla}^{(2)}+\vec{\mathcal
V}(\vec{x}_1,\vec{x}_2)) \Psi(\vec{x}_1,\vec{x}_2)=0
\end{equation}
completing covariantly the eq.(\ref{h}). The compatibility between
eq.(\ref{h}) and the eqs.(\ref{p}) shall be briefly commented in the next
in agreement with the interpretation we give to the interaction potential.

The solution in the coordinate
representation of eqs.(\ref{h}) and (\ref{p}) corresponding to the eigenvalue
$M_{\{n\}}$ where $\{n\}$ is a set of quantum numbers labelling the bound
state is denoted by  $\Psi_{\{n\}}(\vec{x}_1,\vec{x}_2)=\left\langle
\vec{x}_1,\vec{x}_2\vert \Psi_{\{n\}}\right\rangle$.
Its projection on the solutions of the free
Dirac equation $\psi_{\{\vec{k}_j\}}(\vec{x}_j)={\rm exp}(i\epsilon_j\vec{k}_j
\vec{x}_j) w(\{\vec{k}_j\})$ and $\psi^c=
C\bar\psi^T_{\{\vec{k}_j\}}(\vec{x}_j)$ is denoted by $\bar{w}_i(\{\vec{k}_1\})
\left(\Psi_{\{n\}}(\{\vec{k}_1\},\{\vec{k}_2\})\right)_{ij}
\bar{w^c}^T_j(\{\vec{k}_2\})$, where $C$ is the charge conjugation
matrix, $w$ is a Dirac spinor, $\{\vec{k}_j\}$ represents the set of quantum
numbers $\{\epsilon_j,~ r_j,~\vec{k}_j\}$ labelling the free states
($\epsilon_j=\pm$ is the sign of the energy, $r_j$ is the projection of the
spin on an arbitrary axis and $\vec{k}_j$ is the momentum). It is an easy
matter to see that the projection satisfies the following set of equations
derived from (\ref{h}) and (\ref{p}) 

\begin{eqnarray}\label{eqs1}
\left(\epsilon_1\sqrt{\vec{k}_1^2+m_1^2}+\epsilon_2\sqrt{\vec{k}_2^2+m_2^2}
\right)
&\bar{w}&(\{\vec{k}_1\})\Psi_{\{n\}}(\{\vec{k}_1\},\{\vec{k}_2\})
\bar{w^c}^T(\{\vec{k}_2\})+\langle \{\vec{k}_1\},\{\vec{k}_2\} \vert{\mathcal
V}_0\vert\{n\}\rangle\nonumber\\
&=&
M_{\{n\}}~\bar{w}(\{\vec{k}_1\})
\Psi_{\{n\}}(\{\vec{k}_1\},\{\vec{k}_2\})\bar{w^c}^T(\{\vec{k}_2\})
\end{eqnarray}
\begin{equation}\label{eqs2}
(\epsilon_1\vec{k}_1+\epsilon_2\vec{k}_2)
\bar{w}(\{\vec{k}_1\})\Psi_{\{n\}}(\{\vec{k}_1\},\{\vec{k}_2\})
w(\{\vec{k}_2\})+ \langle \{\vec{k}_1\},\{\vec{k}_2\}\vert\vec{\mathcal
V}\vert\{n\}\rangle=0. 
\end{equation}

We now recall that according to the general principles of quantum
mechanics the projection $bar{w}(\{\vec{k}_1\})
\Psi_{\{n\}}(\{\vec{k}_1\},\{\vec{k}_2\})\bar{w^c}^T(\{\vec{k}_2\})$ is the
probability amplitude for finding two free quarks with the individual quantum
numbers $\{\vec{k}_1\}$ and $\{\vec{k}_2\}$ in the meson state characterized
by $\Psi_{\{n\}}$.

Eqs.(\ref{eqs1}) and (\ref{eqs2}) show however that the meson cannot be
represented by a supperposition of states containing free quarks only, because
the sum of the quark 4-momenta does not satisfy the meson mass-shell
constraint. In order to have a real representation, the contribution of the
interaction potential must be explicitly included into the sum, but
the operation must be performed in such a way as to preserve
relativistic covariance.    

The solution we found to this problem was to introduce a classical field
$\Phi$ as an additional component of the meson, independent of the
valence quarks, carrying the 4-momentum $Q^\mu$ \cite{micu} defined as the
difference between the meson and the quark momenta.

\begin{equation}\label{Q}
Q^0=M_{\{n\}}-\epsilon_1\sqrt{\vec{k}_1^2+m_1^2}-
\epsilon_2\sqrt{\vec{k}_2^2+m_2^2} 
\end{equation}
\begin{equation}\label{vQ}
\vec{Q}=-\epsilon_1\vec{k}_1-\epsilon_2\vec{k}_2.
\end{equation}

In this way we can represent a meson 
by a gas of free quarks having the distribution of
momenta and spins given by 
$\bar{w}(\{\vec{k}_1\})\Psi_{\{n\}}(\vec{k}_1,\vec{k}_2)
\bar{w^c}^T(\{\vec{k}_2\}) $ {\it and} a classical effective field which
contributes with its own momentum to the meson. The single meson state is
then:

\begin{eqnarray}\label{meson}
&&\left.\vert {\cal M}_{\{n\}}(M_{\{n\}},0)\right\rangle=
\int d^3k_1~{m_1\over e_1}
d^3k_2{m_2\over e_2}d^4Q 
~\delta^{(3)}(\vec{k}_1+\vec{k}_2+\vec{Q})\delta(e_1+e_2+Q_0-M_{\{n\}})
\nonumber\\
&&\times \sum_{s_1,s_2}
\bar{u}_{s_1}(\vec{k}_1)\Psi_{\{n\}}(\vec{k}_1,\vec{k}_2)C v_{s_2}(\vec{k}_2)
~\Phi^*(Q)~  a^\dagger_{r}
(\vec{k}_1)b^\dagger_{s} (\vec{k}_2)\vert 0\rangle 
\end{eqnarray}
where $\{n\}$ are the quantum numbers of the meson, $a^\dagger$ and
$b^\dagger$ are quark and
antiquark creation operators and $u,v$ are free Dirac
spinors. $\Phi^\dagger$ is the
vacuum-like  classical field carrying the momentum 
$Q^\mu$ defined in (\ref{Q}) and (\ref{vQ}). 
It is the collective, time averaged
effect of the continuous series of virtual excitations of the quark gluonic
field giving rise to the binding. $Q^0$ is the binding energy and $\vec{Q}$
represents the reaction of the whole mass of virtual particles to the motion
of the valence quarks, or, in other words, the effect of the imperfect
cancellation of the vector momenta during the quantum fluctuations. 
One may then say that $\vec{Q}$ is
an aleatory variable defined by eq.(\ref{vQ}) and the eq.(\ref{p}) is an
identity defining the operator $\vec{\mathcal V}(\vec{x}_1,\vec{x}_2)=
i\vec{\nabla}^{(1)}+i\vec{\nabla}^{(2)}$. 
Of course, this not a unique choice for $\vec{\mathcal
V}(\vec{x}_1,\vec{x}_2)$, but it is the simplest one guaranteeing the
compatibility of eqs.(\ref{h}) and (\ref{p}). 

Closing this discussion on the signification of ${\mathrm V}^0$ we notice
a strong similarity with the bag \cite{bm} constraint. Both ${\mathrm V}^0$ and
the bag are classical extra components of the bound system and generate the 
binding effects. We also remark the resemblance between the definition of the
momentum $Q^\mu$ (see eqs.(\ref{Q}),\ref{vQ}) and that of the effective
potential in
nonrelativistic QCD, which is the  remaining part in the effective Lagrangian
after separating off  the kinetic terms \cite{bb,pot}. 

In order to clarify the transformation properties of a meson state
(\ref{meson}) with definite momentum and parity we give the most general
expressions of
$\Psi_{\{n\}}(\{\vec{k}_1\},\{\vec{k}_2\})$ in some simple cases: 

\begin{eqnarray}
&&(\Psi_{P}C)_{lm}=
\varphi_P\gamma^5_{lm}\nonumber\\
&&(\Psi_{V}C)_{lm}=\varepsilon^i\left(
\varphi^V_{1}
\gamma^i_{lm}+\sum_{n=1,2}\varphi^V_{2n}k_n^i\delta_{lm}
+\sum_{n=1,2}\varphi^V_{3n}[\gamma^i,\gamma^j]_{lm}k_n^j
\right)\nonumber\\
&&(\Psi_AC)_{lm}=
\varepsilon^i\left(\varphi^A_{1}\gamma^i_{lp}+
\sum_{n=1,2}\varphi^A_{2n}k_n^i\delta_{lp}
+\sum_{n=1,2}\varphi^A_{3n}[\gamma^i,\gamma^j]_{lp}k^j_n\right)\gamma^5_{pm}
\end{eqnarray}
where $P,~V,~A$ denote the pseudoscalar, vector and axial mesons
respectively, $\varepsilon$ are the meson polarization vectors having only
spatial components in the rest frame and $\varphi_i$ are
scalar functions of $\vec{k}_1,\vec{k}_2$ whose arguments have been omitted for
simplicity. 

It follows then that the expression (\ref{meson}) can be written in a 
Lorentz
covariant form by replacing $\vec{k}$ by $k^\mu_T$ where $k_T=(0,\vec{k})$ in
the rest frame and the scalar product $\vec{k}_i\vec{k}_j$ by  $-k_{iT}^\mu
k_{jT\mu}$. The change of reference frame amounts to the replacement of
$k_T$ by $k^{'\mu}_T=\Lambda(\vec{\omega})^\mu_\nu k^\nu_T$ and of the
Dirac spinors $u,v$ by $u(\vec{k}')=\Lambda_{sp}(\vec{\omega})u(\vec{k})$ where
$\Lambda(\vec{\omega}),~\Lambda_{sp}(\vec{\omega})$ are respectively the
vectorial and spinorial representations of Lorentz transformations from the
rest frame to a reference frame moving with the velocity $\vec{\omega}$ with
respect to the first one. The result is the expression of the meson
state with the energy $E={M\over\sqrt{1-\vec{\omega}^2}}$ and the momentum
$\vec{P}=-\vec{\omega}~E$.

It is worthwhile remarking here that owing to the independent treatment of the
quarks  the generalization of eqs.(\ref{h}), (\ref{p}) and (\ref{meson}) to
the baryon case amounts simply to the introduction of a third quark
contribution according to the general rules of relativistic covariance. 

Closing this discussion we mention that only the projection of $\Psi_{\{n\}}$
on the positive energy states appear in the expression
(\ref{meson}) because those corresponding to negative energy are associated
with the quark and antiquark annihilation operators which gives 0 when acting
on the vacuum. In a forthcoming paper  we shall show that the negative energy
states play an important role in the description of the interaction mechanism
between two hadrons.

Now we verify the normalizability of the single
meson state (\ref{meson}) by  
making use of the commutation relations for the free quark operators and
of a vacuum expectation value of the effective field ensuring the
separate conservation of its 4-momentum:

\begin{equation}\label{vev}
\left\langle 0\right\vert~\Phi(Q)~\Phi^+(Q')~\left\vert 0 
\right\rangle~={1\over V_0T_0}~
\int d^4~X~{\rm e}^{i~(Q'-Q)_\mu~X^\mu}=
{(2\pi)^4\over L^3_0T_0}~\delta^{(4)}(Q-Q')
\end{equation}
where the constant $V_0~T_0$  has been introduced from dimensional reasons.
$V_0$ is the meson volume and $T_0$ is a time sensibly larger than the time
basis involved in the time averaged definition of the effective field.

We notice that 
$\delta(Q_0-Q'_0)$ in (\ref{vev}) induces a cumbersome $\delta(E-E')$
in the expression of the norm. In order to avoid it by preserving the 
manifest Lorentz covariance of eq.(\ref{vev}) we write

\begin{eqnarray}\label{delta} 
{2\pi\over T_0}\delta(Q_0-Q'_0)
&=&{1\over T_0}\int_T dX_0{\rm e}^{i(E(P)-E(P'))X_0}
\approx \nonumber\\
&&{E\over M_{\{n\}}T_0}\int_{T_0}
dY_0 {\rm e}^{i(M_{\{n\}}-M_{\{n'\}})Y_0}\approx{E\over
M_{\{n\}}}\delta_{M_{\{n\}}M_{\{n'\}}} 
\end{eqnarray}
and get immediately:

\begin{equation}\label{norm}
\left\langle~{\cal M}_{\{n'\}}(P')~\vert {\cal
M}_{\{n\}}(P)~\right\rangle~=
2E~(2\pi)^3~\delta^{(3)}(P-P')~\delta_{M_{\{n\}}M_{\{n'\}}}
~\delta_{\{n\}\{n'\}}~{\cal J} 
\end{equation}
where

\begin{eqnarray}\label{J2} 
&&{\cal J}=~{1\over 2M~L_0^3}~\int d^3k_1~{m_1\over
e_1}~d^3k_2~{~m_2\over e_2}~d^4Q~
\delta^{(4)}(k_1+k_2+Q-P)\nonumber\\
&&\times Tr\left({\hat k_1+m_1\over 2m_1}~\Psi_{\{n\}}(k_{1T},k_{2T})
{\hat k_2+m_2\over2m_2}~\bar{\Psi}_{\{n'\}}(k_{1T},k_{2T})\right)=1.
\end{eqnarray}

In the above relations we have implicitly assumed that $M_{\{n\}}$ and
$M_{\{n'\}}$ are discrete eigenvalues of the equation (\ref{h}) with 
$\vert M_{\{n\}}-M_{\{n'\}}\vert~T_0>>1$ so that the integral in (\ref{delta})
vanishes if $M_{\{n\}}\ne M_{\{n'\}}$. We notice that relation (\ref{delta})
allows to eliminate the rather
arbitrary time $T_0$ from the expression of the norm, which is quite 
remarkable. 

As an argument for the introduction of a vector potential in eq. (\ref{p}) we
remark that in its absence i.e. in the case when $Q^\mu=\alpha P^\mu$ the
expression of the  norm will contain the highly singular factor
$\delta(\alpha-\alpha')\delta^{(3)}(0)$. 

Relation (\ref{norm}) can be interpretated as an expression of the
confinement because it shows
that a many particle state (\ref{meson}) can be
normalized like a single particle one if the integral $\cal J$
converges. 

In order to complete the picture we still have to see if the bound state
function $\Psi_{\{n\}}(\vec{x}_1,\vec{x}_2)$ can be recovered from the field
representation (\ref{meson}) of the bound state.  

We mention first that the calculations must be performed in the meson rest
frame, where the bound state wave function has been defined. We also notice 
that there are two kinds of degrees of freedom in the field representation:
those associated with the quarks and those associated with the effective field.
The last ones have no quantum mechanical significance and must be integrated
out. Taking into account the stationarity of the meson structure, its wave
function must be defined as follows:      

\begin{equation}\label{wf}
\tilde\Psi_{\{n\}}(\vec{x}_1,\vec{x}_2,t)_{\alpha \beta}= 
\langle 0\vert\int d^3Q~\bar\psi^c(\vec{x}_2,t)_{\beta}
\bar\psi(\vec{x}_1,t)_{\alpha}\Phi(\vec{Q},t)\vert{\cal
M}_{\{n\}}(M_{\{n\}},0)\rangle  
\end{equation} 
where the single meson state is given by (\ref{meson}), $\psi$ is the free
quark field, $\alpha$ and $\beta$ are spinorial indices and  
\begin{equation}
\Phi(\vec{Q},t)={T_0\over(2\pi)}\int~dQ_0 {\mathrm
e}^{-iQ_0t} \Phi(\vec{Q},Q_0).
\end{equation}

It can be seen by straightforward calculation that the time
dependence factorizes out under the form  ${\rm e}^{-iM_{\{n\}}t}$ (see
eq.(\ref{Q})). It also results that
$\tilde\Psi_{\{n\}}(\vec{x}_1,\vec{x}_2,t)_{\alpha \beta}$ contains that part
of the initial function  $\Psi_{\{n\}}(\vec{x}_1,\vec{x}_2)$ having
nonvanishing projection on the positive energy free states. The part
projecting on the  negative energy free states is lost, because it does not
appear in the definition of the meson state (\ref{meson}). 

In conclusion, we established a relation between the field representation of a
bound state and the solution of the bound state equation in the rest frame. We
have shown that the bound system can be represented by a gas of free particles
and a classical field. The momentum distribution function in the
gas is given by the projection of the bound state wave function on the free
states and the momentum carried by the effective field is
defined as the difference between the bound state momentum and the sum of free
momenta.
We mention that stationarity, Lorentz
covariance, mass-shell constraints and single particle normalizability of the
bound state function are simultaneously and explicitly satisfied. All these
properties have been ensured with the aid of the classical field $\Phi$
representing the collective effect of the unobservable quantum fluctuations
generating the binding. 

The field expression of a bound state (see eq.(\ref{meson})) has clear
transformation properties and can be easily written in any refference
frame. This makes it very useful in
the calculation of some dynamical quantities like, for instance, the
electromagnetic and semileptonic form factors.

\vskip0.5cm     
{\bf Acknowledgments}
This work was started during author's visit at ITP of
the University of Bern in the frame
of the Institutional Partnership Program of the Swiss National
Science  Foundation under Contract No.7 IP 051219. The
author thanks Prof. Heiri Leutwyler for hospitality and stimulating 
discussions.  

A careful reading of the manuscript and suggestions of Floarea
Stancu as well as clarifying discussions with Irinel
Caprini are also gratefully acknowledged.

\end{document}